\documentclass{aa}
\usepackage{graphicx}
\usepackage{txfonts}
\usepackage{psfig}
\usepackage{epsfig}
\usepackage{natbib}
\bibpunct{(}{)}{;}{a}{}{,} 

\begin{document}

\title{The host galaxies of radio-quiet quasars at $0.5<z<1.0$\thanks{Based on observations made with the Nordic Optical Telescope, operated on the island of La Palma jointly by Denmark, Finland,
     Iceland, Norway, and Sweden, in the Spanish Observatorio del 
     Roque de los Muchachos of the Instituto de Astrof{\'\i}sica de 
     Canarias.}}

\author{T. Hyv\"onen\inst{1}, J.K. Kotilainen\inst{1}, E. \"Orndahl\inst{1} \and R. Falomo\inst{2} 
\and M. Uslenghi\inst{3}}

\offprints{T. Hyv\"onen}

\institute{Tuorla Observatory, University of Turku, V\"ais\"al\"antie 20, 
FIN--21500 Piikki\"o, Finland\\
\email{totahy@utu.fi; jarkot@utu.fi; orndahl@gmail.com}
\and
INAF -- Osservatorio Astronomico di Padova, Vicolo dell'Osservatorio 5, I-35122 Padova, Italy\\
\email{falomo@pd.astro.it}
\and
INAF -- IASF Milano, Via E. Bassini 15, I-20133 Milano, Italy\\
\email{uslenghi@mi.iasf.cnr.it}
}

\date{Received; accepted }

\abstract{
We present near-infrared $H$-band imaging of 15 intermediate redshift ($0.5<z<1$) 
radio quiet quasars (RQQ) in order to characterize the properties of their host galaxies. 
We are able to clearly detect the surrounding nebulosity in 12 objects, 
whereas the object remains unresolved in three cases. For all the resolved objects, 
we find that the host galaxy is well represented by a de Vaucouleurs $r^{1/4}$ surface brightness law. 
This is the first reasonably sized sample of intermediate redshift RQQs studied 
in the near-infrared.

The RQQ host galaxies are luminous (average $M_H=-26.3\pm0.6$) 
and large giant elliptical galaxies (average bulge scale length $R_e = 11.3\pm5.8$ kpc). 
RQQ hosts are $\sim1$ mag brighter than the typical low redshift galaxy luminosity $L^*$, and 
their sizes are similar to those of galaxies hosting lower redshift RQQs, indicating that there is 
no significant evolution at least up to z$\sim$1 of the host galaxy structure. 
We also find that RQQ hosts are $\sim0.5-1$ mag fainter than 
radio-loud quasars (RLQ) hosts at the similar redshift range.  
The comparison of the host luminosity of intermediate redshift RQQ hosts with that for lower  $z$ sources 
shows a trend that is consistent with that expected from the passive evolution 
of the stars in the host galaxies.

The nuclear luminosity and the nucleus/host galaxy luminosity ratio of the objects 
in our sample are intermediate between  those of lower redshift RQQs and those of  
higher redshift ($z>1$) RQQs. 


}

\titlerunning{Evolution of RQQ hosts}
\authorrunning{Hyv\"onen et al.}

\maketitle

%

\defcitealias{koti98a}{K98}
\defcitealias{koti00}{K00}

\section{Introduction}

To understand the quasar phenomenon, and AGN in general, it is 
important to study their orientation--independent properties, such as the 
luminosity of their host galaxies and their large scale environment. 
The relationship between quasars and their host galaxies, in particular, is a key ingredient for 
understanding the quasar activity, the formation of galaxies and the strong 
cosmological evolution of the space density of the quasar population \citep{warr94}. 
The shape of this evolution as a function of $z$ is similar to 
that of the black hole (BH) mass accretion rate and 
the cosmic star formation history of the Universe \citep{mada98,fran99,char01,barg01,yu02,marc04}, 
suggesting a fundamental relationship between the processes of formation of 
massive galaxy bulges and their nuclei. Thus, studies of quasars and their 
host galaxies will give us a more detailed understanding of galaxy evolution. 
According to the hierarchical model of structure formation \citep{kauff00,dima03} 
there should be a correlation between the evolution of massive spheroids and the 
processes that fuel their central BHs.

Inactive supermassive BHs are prevalent in nearby inactive 
massive spheroids \citep[e.g.][]{ferr02,bart04}. 
Recently, at low redshift, correlations have been found between the 
BH mass, and the luminosity and the central stellar velocity dispersion 
of the host galaxy bulge \citep[e.g.][]{mago98,vmar99,gebh00,ferr00,mclu02,marc03,bett03,hari04}. 
These findings suggest that nuclear activity is a common phenomenon during 
the lifetime of a massive galaxy with recurrent accretion episodes, and that the nuclear power depends 
on the mass of the galaxy.

Imaging of quasar host galaxies from space (e.g. HST) has the advantage 
over ground-based observations of having an excellent spatial resolution (narrow PSF) 
to resolve the host galaxy close to the nucleus. However, HST has a relatively 
small aperture that translates into a limited capability to detect faint extended 
nebulosity. A number of ground-based \citep[e.g.][]{mcle94a,mcle94b,perc01} and space-based 
\citep[e.g.][]{bahc97,hami02,dunl03,paga03,floy04} optical and near-infrared (NIR) studies 
of low redshift quasar hosts ($z<0.5$) have shown that both radio-quiet (RQQ) 
and radio-loud (RLQ) quasars are predominantly hosted 
by luminous, massive elliptical galaxies that are typically brighter than 
$L^*$ galaxies (the Schechter function's characteristic luminosity) and as bright as 
low redshift radio galaxies (RG). Some of these studies also suggest that 
the morphology of the host may depend on the nuclear luminosity in the sense that 
high luminosity RQQs are exclusively hosted in elliptical galaxies 
(or dominated by the spheroidal component) 
while low luminosity RQQ can also be located in early-type spiral (disc-dominated) galaxies 
\citep{hami02,dunl03}.

To investigate the evolution of quasar host galaxies as a function of 
cosmic time, observations are needed over a large redshift range. Unfortunately, 
high redshift ($z>0.5$) hosts have been until recently relatively little studied because of 
difficulties due to limited spatial resolution and the high nucleus/host luminosity 
contrast (due to $(1+z)^4$ cosmological surface brightness dimming). 
In fact, although a number of detections of quasar hosts at z $>$ 1 based on ground-based 
\citep[e.g.][]{falo01,falo04,falo06,koti06} and space observations \citep[e.g.][]{kuku01,peng06} are 
available, the only systematic observations of large samples at 
intermediate redshift ($0.5<z<1$) in the NIR have been made by \citet{koti98a} and 
\citet{koti00} (hereafter \citetalias{koti98a} and 
\citetalias{koti00}, respectively) for a total sample of $\sim40$ flat and steep spectrum RLQs 
(FSRQ and SSRQ, respectively). 
In particular, no homogenous studies of matched samples of RLQs and RQQs at intermediate redshift 
have yet been made. Such a sample would allow to study and compare the evolution of the hosts of 
both types of quasars. 

The classes of RLQs and RQQs differ in their radio properties but they also appear to be different 
in their host galaxy luminosity \citep[][\citetalias{koti98a,koti00}]{dunl03,floy04,falo04}, 
in the sense that RLQ hosts are significantly more luminous than RQQ hosts at all redshifts in 
the redshift range $0<z<2$. In addition, both types of quasars follow the passive evolution of 
massive elliptical galaxies. This is inconsistent with models of hierarchical galaxy 
formation \citep{kauff00,dima03}, which predict a decrease in the luminosity (mass) of the galaxies 
at high redshift, followed by an increase toward the present epoch due to galactic mergers.

\begin{figure}
\centering
\resizebox{8cm}{8cm}{\includegraphics{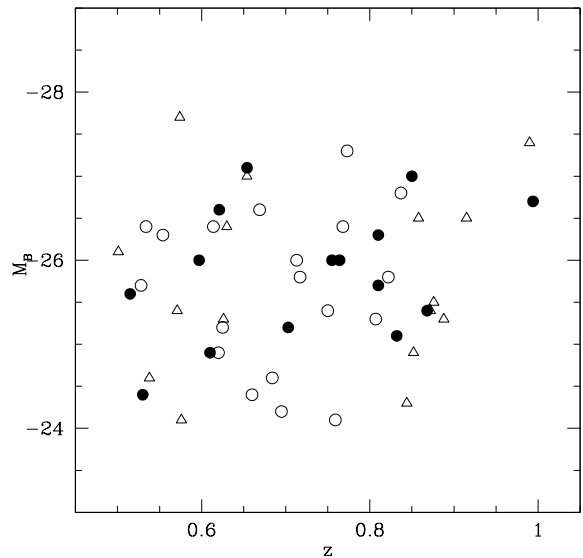}}
\caption{The absolute $B$-band magnitude \citep{vero03} of the intermediate redshift RQQs and RLQs 
versus redshift. RQQs from this work are marked as filled circles, SSRQs from \citetalias{koti00} as 
open circles and FSRQs from \citetalias{koti98a} as open triangles.
\label{MBz}}
\end{figure}

\setlength{\tabcolsep}{0.8mm}
\begin{table}
\caption{The sample and the journal of observations.$^{\mathrm{a}}$}
\label{journal}
\begin{tabular}{@{}lllllcrl@{}}
\hline
\noalign{\smallskip}
Name & $z$ & $V$ & $M_B$ & Date & T$_{exp}$ & Seeing & $\varepsilon$ \\
& & & & & min & arcsec &\\
(1) & (2) & (3) & (4) & (5) & (6) & (7) & (8) \\
\noalign{\smallskip}
\hline
\noalign{\smallskip}
HS 0010+3611   & 0.530  & 18.2 & $-$24.4 & 27/08/2004 & 45  & 0.80 & 0.05\\
1E 0112+3256   & 0.764  & 18.9 & $-$26.0 & 27/08/2004 & 55  & 0.80 & 0.22\\
PB 6708        & 0.868  & 18.6 & $-$26.6 & 24/01/2005 & 64  & 0.95 & 0.08\\
KUV 03086-0447 & 0.755  & 17.5 & $-$25.7 & 23/01/2005 & 106 & 0.95 & 0.03\\
US 3828        & 0.515  & 16.9 & $-$25.1 & 22/01/2005 & 92  & 0.95 & 0.06\\
MS 08287+6614  & 0.610  & 18.0 & $-$26.3 & 23/01/2005 & 73  & 0.85 & 0.10\\
TON 392        & 0.654  & 16.0 & $-$25.6 & 24/01/2005 & 44  & 1.00 & 0.11\\
US 971         & 0.703  & 18.1 & $-$27.1 & 22/01/2005 & 61  & 0.80 & 0.07\\
HE 0955-0009   & 0.597  & 16.9 & $-$25.4 & 23/01/2005 & 104 & 0.90 & 0.04\\
HE 1100-1109   & 0.994  & 17.5 & $-$25.2 & 22/01/2005 & 73  & 0.90 & 0.10\\
CSO 769        & 0.850  & 16.9 & $-$24.4 & 23/01/2005 & 53  & 0.85 & 0.15\\
HS 1623+7313   & 0.621  & 16.3 & $-$27.0 & 28/08/2004 & 75  & 0.85 & 0.04\\
HS 2138+1313   & 0.810  & 18.1 & $-$24.9 & 28/08/2004 & 60  & 1.00 & 0.16\\
LBQS 2249-0154 & 0.832  & 18.7 & $-$26.0 & 28/08/2004 & 64  & 0.90 & 0.10\\
ZC 2351+010B   & 0.810  & 17.5 & $-$26.7 & 28/08/2004 & 40  & 1.00 & 0.16\\
\noalign{\smallskip}
\hline
\end{tabular}
\begin{list}{}{}
\item[$^{\mathrm{a}}$] 
Column (1) gives the name of the object; 
(2) the redshift; (3) the $V$-band apparent magnitude; (4) the $B$-band absolute 
magnitude; (5) the date of observation; (6) total exposure time; (7) seeing FWHM (arcsec); and 
(8) the average ellipticity of the field stars.
\end{list}
\end{table}
\setlength{\tabcolsep}{1.5mm}

We present here a deep high spatial resolution 
NIR $H$-band imaging study of the host galaxies of a sizeable sample of RQQs at intermediate redshift. 
For all the RQQs in the sample these are the first high quality NIR imaging observations, and the 
first detection of the host galaxies. 
The RQQ sample was extracted from the quasar catalog of \citet{vero03}, as having $0.5<z<1$, $-24<M_B<-28$, 
and no radio emission listed in that catalog or in the catalog of the FIRST 1.4 GHz survey \citep{firs97}. 
All selected targets were also required to have at least one 
sufficiently bright field star within 2 arcmin distance from the quasar, 
to allow a reliable characterization of the PSF. In total, 15 RQQs were observed. 
Fig.~\ref{MBz} shows their distribution in the $z-M_B$ plane, compared to the FSRQ and SSRQ samples 
\citepalias{koti98a,koti00}. 
The average redshift and the absolute magnitude of the RQQ sample are $z=0.728\pm0.138$ and 
$M_B=-25.8\pm0.9$, respectively. For the combined sample of FSRQs and SSRQs from \citetalias{koti98a} and 
\citetalias{koti00} the corresponding average values are $z=0.719\pm0.138$ and $M_B=-25.7\pm1.0$. 
The RQQ sample is thus well matched with the FSRQs and SSRQs samples in both redshift and absolute magnitude, 
allowing us to homogeneously compare the properties of RQQs and RLQs in this redshift range. 
The general properties of the observed objects are given in Table~\ref{journal}.

The immediate objective of this study is to resolve the host galaxies of the quasars 
and to derive their global properties: e.g. 
absolute magnitude, effective radius, central surface brightness and the nucleus/host galaxy 
luminosity ratio. These properties will allow us to assess the luminosity gap between 
the RLQ and RQQ hosts at intermediate redshift and to investigate in more detail the 
cosmological evolution of both types of quasars between the peak of the quasar activity 
($z\sim2-3$) and the present epoch. The determination of the RQQ host properties 
in the redshift gap $0.5<z<1$ is an important contribution for the study of 
the behaviour of the RQQ host luminosity as a function of cosmic time. At lower redshift, 
volume selection effects are more severe and current 
samples of resolved RQQ are rather poorly defined. In addition, we shall investigate the dependence of the 
host galaxy properties on the nuclear luminosity in our quasar sample and also compare 
the occurence of close companions between RLQs and RQQs, and compared to lower and higher redshift, 
in order to assess the importance of interactions for the triggering and fueling of quasar activity.

The outline of the paper is as follows. In Section 2, we describe 
the observations, data reduction and the method of the analysis. In Section 3, 
we present the results and discussion, concerning the host galaxy properties, the nuclear component 
and the close environment of the RQQs. Finally, summary and main conclusions are given in Section 4. 
Throughout this paper, to facilitate comparison with the FSRQs and SSRQs, 
$H_0=50$ kms$^{-1}$Mpc$^{-1}$ and $q_0=0$ cosmology is used. The 
conclusions do not change significantly if the concordant $\Lambda$-cosmology is used.

\section{Observations, data reduction and analysis}

The observations were carried out during two observing runs in August 2004 and January 2005 
at the 2.5m Nordic Optical Telescope (NOT). We used the $1024\times1024$~pixel NOTCam NIR 
detector with a pixel scale of 0\farcs235~pixel$^{-1}$, giving a field of view of 
$\sim4\times4$ arcmin$^2$. The $H$-band ($1.65~\mu$m), corresponding to 
$\sim0.8-1.1~\mu$m rest-frame wavelength, was used for all the observations. 
This filter choice minimizes the nucleus/host brightness contrast, while the extinction, the $K$-correction 
and the contribution from quasar emission lines and scattered nuclear light are insignificant. In addition, 
these observations can be directly compared with our previous FSRQ and SSRQ data. 
The seeing during the observations, as derived from the median FWHM size of the stars in each frame, 
ranged from 0\farcs75 to 1\farcs00 arcsec FWHM (mean and median 
0\farcs9 FWHM). A journal of the observations is given in Table~\ref{journal}.
The images were acquired by dithering the targets across the array in a random grid 
within a box of about 20 arcsec, and taking a 60 sec exposure at each position, 
always keeping the target well inside the field. Individual exposures were then coadded to 
achieve the final integration time for each object (Table~\ref{journal}).

\subsection{Data reduction}

Data reduction was performed using IRAF\footnote{IRAF is distributed by 
the National Optical Astronomy Observatories, which are operated by 
the Association of Universities for Research in Astronomy, Inc., 
under cooperative agreement with the National Science Foundation.} 
and closely follows the method presented in \citet{koti05}. Bad pixels were 
corrected for in each image using a mask made from the ratio of two sky flats with different 
illumination levels. Sky subtraction was performed 
for each science image using a median averaged frame 
of all the other temporally close frames in a grid of eight exposures. 
Flat fielding was made using normalized median averaged twilight sky frames 
with different illumination levels. Finally, images of the same target were aligned to sub-pixel accuracy 
using field stars as reference points and combined after removing spurious pixel values 
to obtain the final reduced co-added image. Standard stars from \citet{hunt98} were 
observed throughout the nights in order to provide 
photometric calibration, which resulted in internal photometric accuracy of $\sim0.1$~mag 
as determined from the comparison of all observed standard stars. 

\subsection{2D Analysis}

Two-dimensional analysis was carried out using AIDA 
(Astronomical Image Decomposition and Analysis, Uslenghi \& Falomo, in prep.), 
a software package specifically designed to perform two-dimensional
model fitting of quasar images, providing simultaneous decomposition
into nuclear and host components. The analysis consists of two main parts:
a) PSF modeling and b) quasar host characterization.

\subsubsection{PSF modeling}

To detect the host galaxies of quasars and to characterize their properties, 
the key factors are the nucleus-to-host luminosity ratio and the seeing (the shape of the PSF). 
The most critical part of the analysis is thus to perform a detailed study of the PSF for each frame. 
The PSF modeling is based on fitting a parameterized bidimensional model to the stars.

For PSF modeling, the point-like objects of the image were classified as stars based on their FWHM, 
sharpness and roundness. When possible, a sufficiently bright, saturated star was included 
in the list of reference stars in order to model the shape of the PSF wings, against which most of the signal 
from the surrounding nebulosity will be detected. The relatively large field of view of NOTCam 
($\sim4\times4$ arcmin$^2$) and the constraint on the quasar selection to have at least one bright star 
within 2 arcmin from the quasar, 
allowed us to reach this goal and thus to perform a reliable characterization of the PSF. 
Images with a large number of stars distributed over the field of view have been checked to account for 
any possible positional dependence of the PSF. No significant variations were found and in this analysis 
the PSF has been assumed to be spatially invariant, i.e., the same model has been fitted simultaneously to 
all the reference stars of the image.

For each star, a mask was created to exclude contamination from 
nearby sources, bad pixels and other defects affecting the image. 
Then the local background was computed in a circular annulus 
centered on the source, with the MMM algorithm (from IDL Astrolib). 
Its uncertainty was estimated from the standard deviation of the 
values computed in sectors of concentric sub-annuli included in this area. 
The region to be used in the fit was selected by defining an inner and outer radius of a circular area. 
The inner radius was set to zero, except for bright saturated stars, thus allowing their core to be 
excluded from the fitting. The outer radius was generally set to correspond to a distance where 
the signal in the radial profile was not significantly higher than the noise.


Stellar images were found to be slightly elliptical, with FWHM varying from 0.80 to 1.05 arcsec and 
average ellipticity at half maximum $\varepsilon\sim0.15$. 
Characteristic average values for the stars in each quasar image are presented in Table~\ref{journal}. 
As stated above, the goodness of the model is strictly related to the availability of suitable stars 
in the images. Only six images contain a bright enough star to model the PSF wings at 
radius $\geq3$ arcsec. However, for all the images we were able to extract a PSF model up to 
the maximum radius where the signal from the quasar was present. 
The worst case in this respect is ZC2351+010B, with only one star available due to 
the highly irregular background and the presence of image defects which prevented the use of 
the only bright star in the field.

\subsubsection{Quasar host characterization}

Once a suitable representation of the PSF was determined, the quasar images
(prepared in an identical way to that described for the stars in the previous section) 
were first fitted with only the PSF model in order to provide a first order indication 
of a deviation from the PSF shape. If the residuals did not reveal any systematic excess, 
the object was considered unresolved. Otherwise, the object was fitted with a scaled PSF to represent 
the nucleus, and a de Vaucouleurs r$^{1/4}$ model or an exponential disk model convolved with the PSF 
to represent the host galaxy, using an iterative least-squares fit to the observed data. 

With the applied procedure we have derived the luminosity and the scale-length 
of the host galaxies and the luminosity of the nuclei. 
An estimate of the errors associated with the computed parameters was obtained by Monte Carlo 
simulation of synthetic data sets. Simulated quasar images were generated adding noise to 
the best fit model, then the fit procedure was applied to these images, 
producing a "best fit" combination of parameters for each image. 
For each parameter, the standard deviation of the best-fit values gives an estimate of 
the uncertainty on the parameters under the assumption that the distribution of the 
errors does not vary rapidly as a function of the values of the parameters. 
Obviously, this procedure does not take into account 
systematic errors generated by an imperfect modelling of the PSF, which can be roughly estimated
by comparing the results obtained with different PSF models, statistically consistent with 
the available data. For our worst case, ZC2351+0108, for example, this effect produces 
an uncertainty of $\sim0.6$ magnitude for the brightness of the host galaxy. 
Instead, in images that have several suitable reference stars (e.g. TON 392), 
the uncertainty is dominated by the noise.

Upper limits to the host magnitudes for unresolved objects were computed by adding 
a galaxy component to the PSF and varying its surface brightness until the model profile 
was no longer consistent with the object profile.

While the total magnitude of the host galaxy can be derived with a typical internal error of 
0.2 - 0.6 mag (0.3 mag on average), the scale-length is often poorly constrained. 
This depends on the degeneracy that occurs between the effective radius $r_e$ and 
the surface brightness $\mu_e$. Estimated errors of the effective radius and host galaxy magnitude for each 
target are given in Table~\ref{hostprop}. No correlation is apparent between $r_e$ and 
seeing or nuclear magnitude.

At these relatively high redshifts, it is difficult to distinguish between the exponential disk 
and the bulge (r$^{1/4}$ law) models for the host galaxy from the luminosity distributions. 
However, in all cases a bulge model resulted in a better fit (lower $\chi^{2}$ value) than 
a disk model, and in what follows, we have assumed that the host galaxies can be represented as 
elliptical galaxies following a de Vaucouleurs model. This is supported by the strong evidence at 
low redshift for the predominance of bulge dominated hosts of quasars \citep{hami02,dunl03,paga03}. 
Note that adopting instead a disk model would result in fainter host galaxies, 
by $\sim$0.5 mag on average, but this would introduce no systematic differences to our conclusions.

\section{Results and discussion}



\begin{figure*}
\centering
\includegraphics{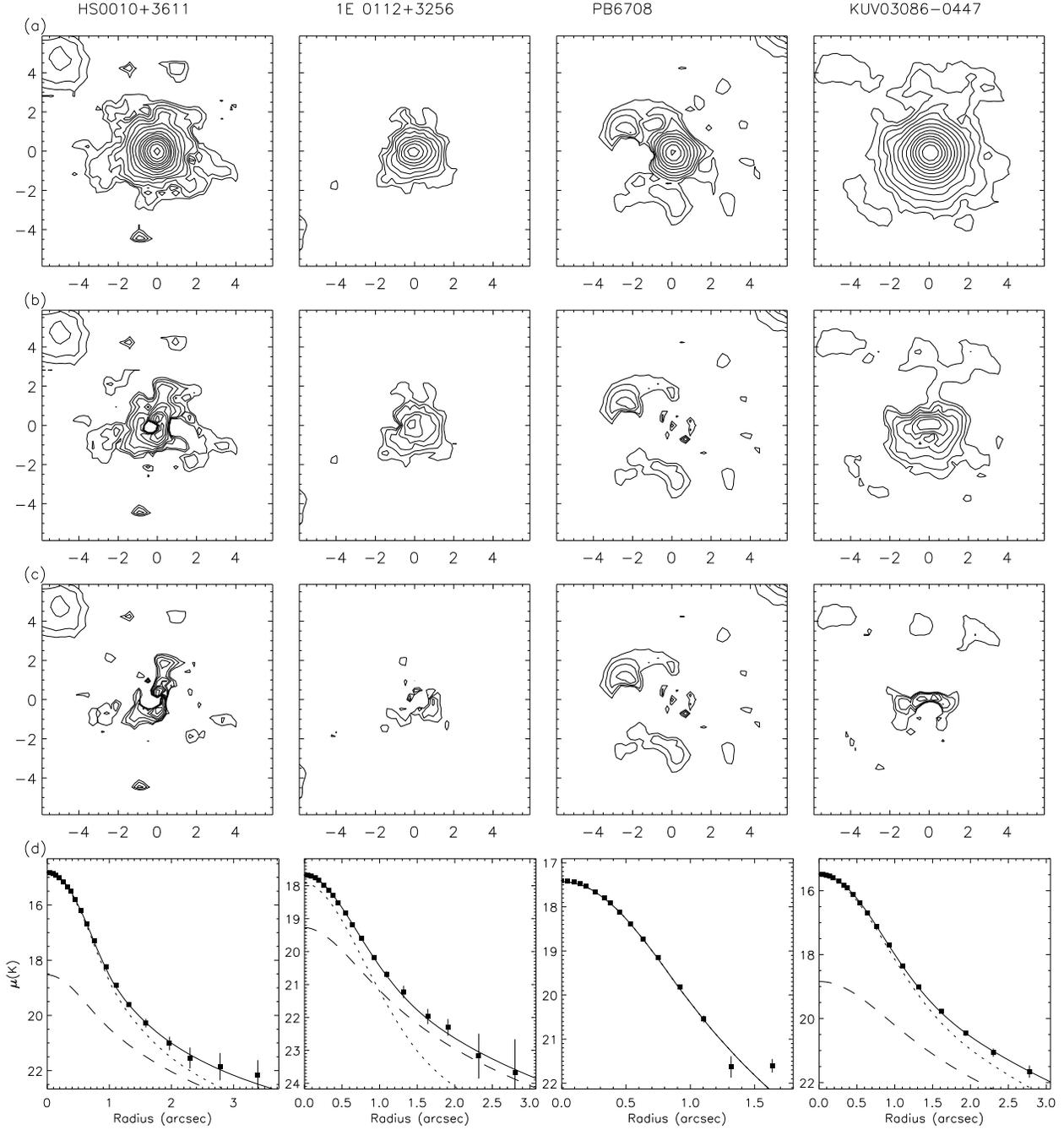}
\caption{The $H$-band images of the quasars, from top to bottom a) the original image 
b) the PSF model subtracted image of the host galaxy and c) residuals after the model fitting. Panel d) shows 
the observed radial surface brightness profiles (solid points with error bars) for each 
quasar, overlaid with the scaled PSF model (dotted line), the de Vaucouleurs $r^{1/4}$ model 
convolved with the PSF (long-dashed line) and the fitted PSF + host galaxy model profile (solid line). 
The Y-axis is in mag~arcsec$^{-2}$.
\label{radial}}
\end{figure*}

\setcounter{figure}{1}
\begin{figure*}
\centering
\includegraphics{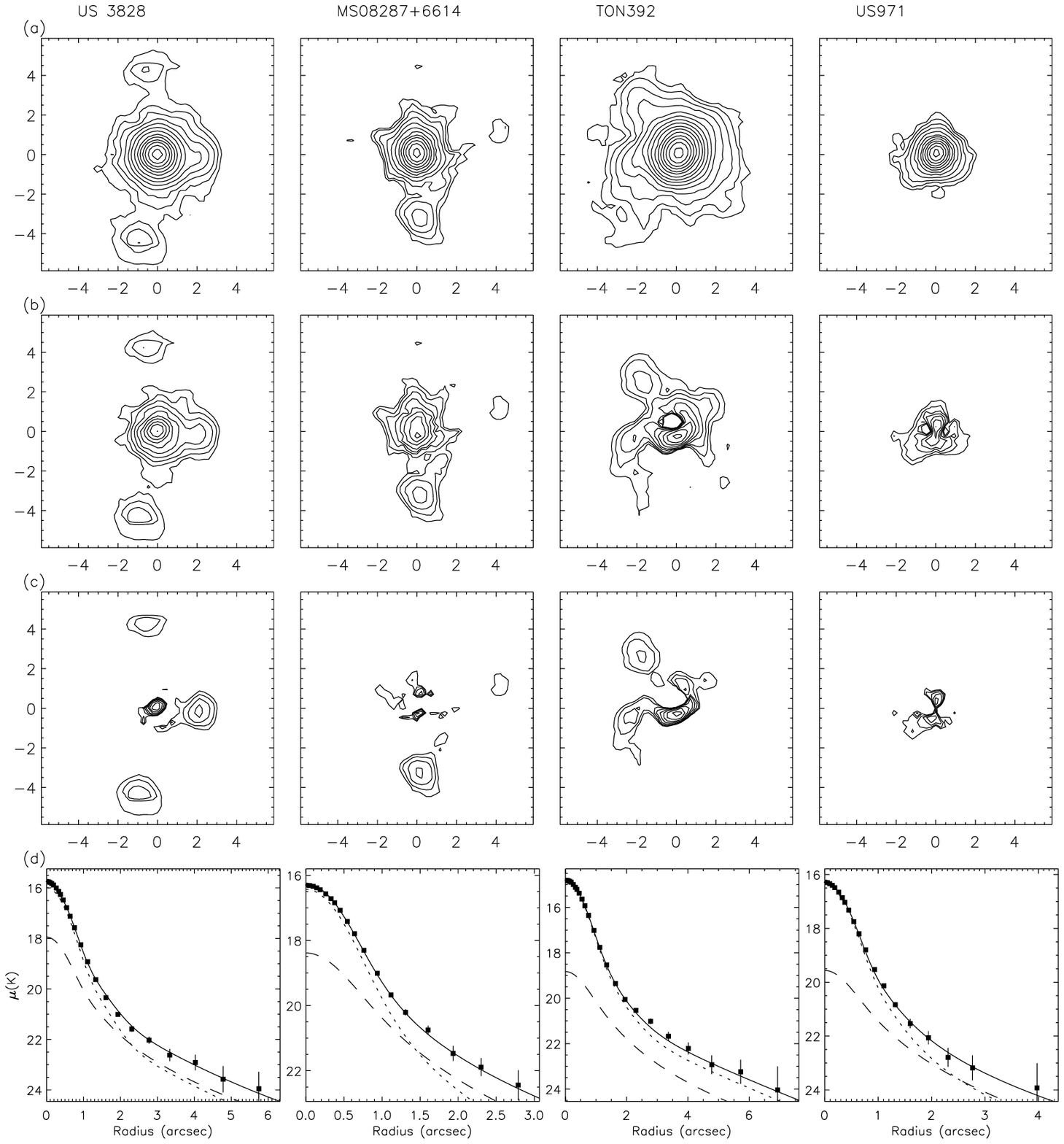}
\caption{(continued)}
\end{figure*}

\setcounter{figure}{1}
\begin{figure*}
\centering
\includegraphics{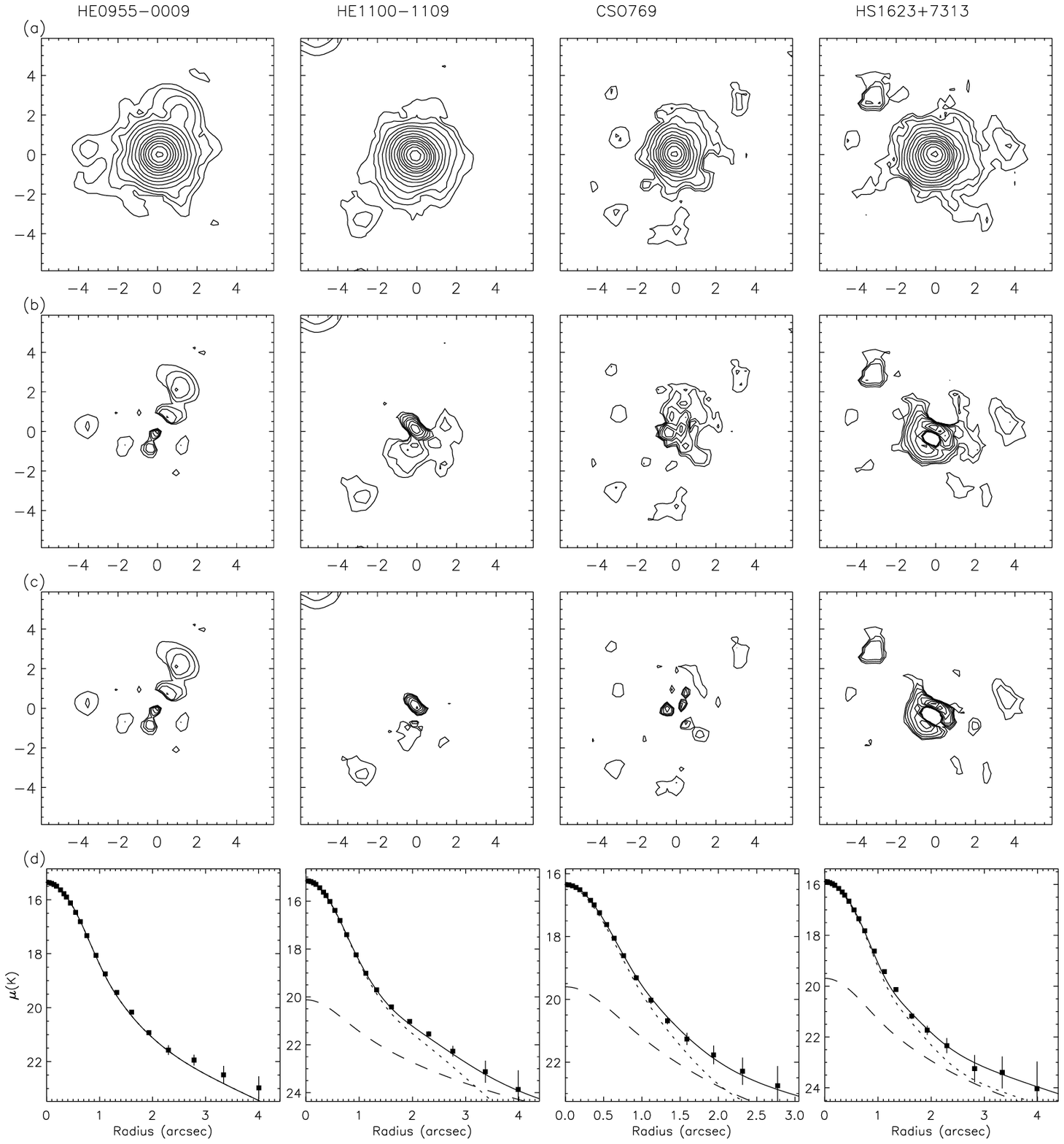}
\caption{(continued)}
\end{figure*}

\setcounter{figure}{1}
\begin{figure*}
\centering
\includegraphics{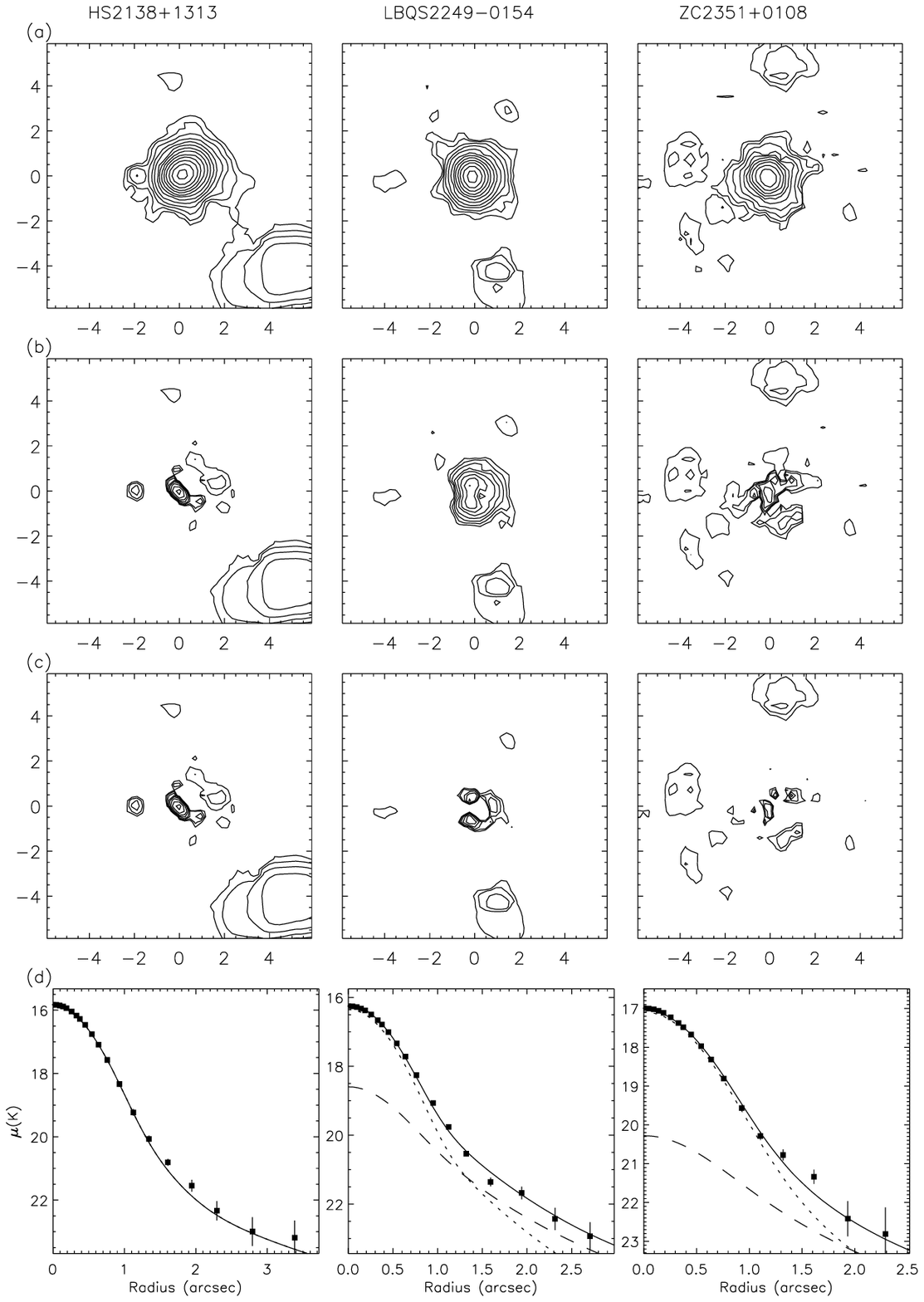}
\caption{(continued)}
\end{figure*}

\begin{table*}
\centering
\caption{Properties of the host galaxies.$^{\mathrm{a}}$}
\label{hostprop}
\begin{tabular}{llllllllllllll}
\hline
\noalign{\smallskip}
Name & $z$ & $K-$corr & $m_{H, nuc}$ & $m_{H, host}$ &$\mu_e^{\mathrm{b}}$ & $r_e$ & $R_e$ & $M_{H, nuc}^{\mathrm{c}}$ & $M_{H, host}^{\mathrm{d}}$ & $L_{nuc}/L_{host}$ & Note & $\chi^2_{psf+gal}/\chi^2_{psf}$\\
& & mag & mag & mag & mag & arcsec & kpc & mag & mag & & & &\\
(1) & (2) & (3) & (4) & (5) & (6) & (7) & (8) & (9) & (10) & (11) & (12) & (13) \\
\noalign{\smallskip}
\hline
\noalign{\smallskip}
HS 0010+3611   & 0.530 & 0.12 & 15.03 & 17.26    & 14.68 & 1.3$\pm0.6$ & 10.5$\pm5.0$ & $-$28.0 & $-$25.9$\pm0.3$  & 7.8    & R & 0.93\\
1E 0112+3256   & 0.764 & 0.18 & 17.94 & 18.18    & 15.46 & 0.9$\pm0.5$ & 8.9$\pm4.9$ & $-$26.1 & $-$26.0$\pm0.3$  & 1.25    & R & 0.70\\
PB 6708        & 0.868 & 0.22 & 17.13 & $>$18.50 &       &      &      & $-$27.2 & $>$$-$26.1       & $>$3.5 & U &     \\
KUV 03086-0447 & 0.755 & 0.17 & 15.16 & 17.14    & 14.61 & 1.0$\pm0.5$ &  9.7$\pm4.9$ & $-$28.8 & $-$27.0$\pm0.3$  & 6.2    & R & 0.63\\
US 3828        & 0.515 & 0.11 & 15.70 & 17.28    & 14.52 & 1.2$\pm0.5$ &  9.8$\pm4.1$ & $-$27.3 & $-$25.8$\pm0.3$  & 6.9    & R & 0.85\\
MS 08287+6614  & 0.610 & 0.13 & 16.45 & 17.42    & 13.97 & 0.8$\pm0.4$ &  6.9$\pm3.6$ & $-$27.0 & $-$26.1$\pm0.2$  & 2.4    & R & 0.65\\
TON 392        & 0.654 & 0.14 & 14.41 & 17.07    & 14.23 & 1.0$\pm0.4$ &  8.9$\pm3.7$ & $-$29.2 & $-$26.6$\pm0.3$  & 11.6   & R & 0.93\\
US 971         & 0.703 & 0.15 & 16.57 & 18.27    & 15.78 & 1.1$\pm0.6$ & 10.1$\pm5.7$ & $-$27.2 & $-$25.7$\pm0.2$  & 4.8    & R & 0.94\\
HE 0955-0009   & 0.597 & 0.13 & 15.18 & $>$17.20 &       &      &      & $-$28.2 & $>$$-$26.3       & $>$6.4& U &     \\
HE 1100-1109   & 0.994 & 0.30 & 15.23 & 17.93    & 17.99 & 2.5$\pm0.7$ & 27.3$\pm7.6$ & $-$29.5 & $-$27.1$\pm0.2$  & 12.0    & R & 0.86\\
CSO 769        & 0.850 & 0.22 & 16.42 & 17.99    & 16.19 & 1.2$\pm0.5$ & 13.4$\pm5.1$ & $-$27.9 & $-$26.5$\pm0.3$  & 4.2    & R & 0.98\\
HS 1623+7313   & 0.621 & 0.14 & 15.86 & 18.18    & 15.68 & 1.2$\pm0.5$ & 10.5$\pm4.5$ & $-$27.6 & $-$25.4$\pm0.2$  & 8.5    & R & 0.86\\
HS 2138+1313   & 0.810 & 0.20 & 15.57 & $>$18.00 &       &      &      & $-$28.6 & $>$$-$26.4       & $>$9.4 & U &     \\
LBQS 2249-0154 & 0.832 & 0.21 & 16.38 & 17.43    & 14.36 & 0.7$\pm0.3$ &  7.1$\pm3.1$ & $-$27.9 & $-$27.0$\pm0.5$  & 2.6   & R & 0.67\\
ZC 2351+010B   & 0.810 & 0.20 & 16.86 & 18.52    & 16.59 & 1.2$\pm0.6$ & 12.3$\pm6.1$ & $-$27.3 & $-$25.9$\pm0.6$  & 4.6    & R & 0.92\\
\noalign{\smallskip}
\hline
\end{tabular}
\begin{list}{}{}
\item[$^{\mathrm{a}}$]
Columns (1) and (2) give the name and the redshift of the object; 
(3) the $K-$correction for 
first-ranked elliptical galaxies from \citet{neug85}, 
interpolated to the redshifts of the RQQs; (4) and (5) the apparent $H$-band nuclear 
and host galaxy magnitude; (6) the effective surface brightness $\mu_e$; 
(7) and (8) the bulge scale-length in arcsec and kpc; 
(9) and (10) the absolute $H$-band nuclear and host galaxy magnitude; 
(11) the nucleus/host galaxy luminosity ratio; (12) R = resolved; 
U = unresolved; and (13) the ratio of the $\chi^2_{psf+gal}$ and $\chi^2_{psf}$ . 
\item[$^{\mathrm{b}}$] 
Corrected for cosmological dimming.
\item[$^{\mathrm{c}}$] 
The RQQ nuclei are assumed to have power-law spectra with $\alpha\sim-1$ 
and therefore to have negligible $K-$correction.
\item[$^{\mathrm{d}}$] 
Corrected for $K-$correction.
\end{list}
\end{table*}

In Fig.~\ref{radial} we show $H$-band contour plots for the quasars, the PSF model subtracted image of 
the host galaxy, the residuals after the model fitting and the radial surface brightness profiles with 
the best fit model from the procedure described in the previous Section. 
We were able to find significant deviations of the radial profile with respect to its PSF, and thus 
clearly detect the host galaxy in 12 objects, whereas the host galaxy remained unresolved in three objects. 
The nuclear and host galaxy magnitudes, and the effective radii of the host galaxies are summarized 
in Table~\ref{hostprop}.
The absolute magnitudes of the host galaxies have been $K-$corrected using 
the optical-NIR evolutionary synthesis model for elliptical galaxies \cite{neug85}. 
The size of this correction is $m_H\sim0.16$ at the average redshift of the RQQ sample, 
$z\sim0.73$ (see Table~\ref{hostprop}). No $K-$correction was applied to the nuclear component, 
which was assumed to have a power-law spectrum ($f_\nu \propto \nu^{-\alpha}$) with $\alpha\sim-1$. 
No correction for Galactic extinction was applied since it is negligible in the observed $H$-band.

Because of the three upper limits in our sample, we used the statistical 
survival analysis method for right censored data \citep{feig85} 
to estimate the mean value $\mu$ of the luminosity of the host galaxies. 
We used 
\[\mu=\sum_{j=1}^{r+1}S[x_j][x_j-x_{j+1}]\] 
where $S[x_j]$ is the Kaplan--Meier estimator, $x_j$ is each value 
of the sample and $x_r$ is the maximum value of the sample. 
The variance was also estimated using the method in \citet{feig85} but its contribution to the error is 
insignificant. In all cases, the upper limits to the host magnitudes are 
fainter than $M_H=-26.4$. If the mean value is calculated using only the resolved hosts, we 
are biased against the faint part of the population and slightly overestimate the luminosity.
 

\subsection{Luminosities and sizes of the host galaxies}

\setlength{\tabcolsep}{3.5mm}
\begin{table*}
\centering
\caption{Comparison of average host galaxy properties with other RQQ samples.$^{\mathrm{a}}$}
\label{samples}
\begin{tabular}{lllllll}
\hline
\noalign{\smallskip}
Sample & filter & $N$ & $<z>$ & $M_B$ & $<M_{H, nuc}>$$^{\mathrm{c}}$ & $<M_{H, host}>$$^{\mathrm{c}}$\\
(1) & (2) & (3) & (4) & (5) & (6) & (7)\\
\noalign{\smallskip}
\hline
\noalign{\smallskip}
RQQ (this work)    & $H$ & 12  & 0.720$\pm$0.142 & $-25.8\pm0.9$ &$-$27.8$\pm$1.1 & $-$26.3$\pm$0.6\\
RQQ/R+U$^{\mathrm{b}}$ (this work) & $H$ & 15 & 0.728$\pm$0.138 &  $-25.8\pm0.9$ &$-$27.9$\pm$1.0 & $-$26.2$\pm$0.6\\
                     &      &       &                &                   &            \\
RQQ \citep{mcle94a} & $H$ & 22/24 & 0.103$\pm$0.029 &              &$-$25.1$\pm0.5$  &$-$24.9$\pm$0.6\\
RQQ \citep{tayl96} & $K$ & 19/19 & 0.157$\pm$0.062 & $-$23.8$\pm$0.6 &$-$26.1$\pm0.9$ &$-$25.7$\pm$0.7\\
RQQ \citep{dunl03}  & $R$ & 13/13 & 0.175$\pm$0.01 &                 & $-$24.3$\pm$0.6 &$-$25.9$\pm$0.2 \\
RQQ \citep{bahc97} & $V$ & 14/14 & 0.183$\pm$0.046 & $-24.9\pm0.5$ & &$-$25.1$\pm$0.6 \\
RQQ \citep{mcle94b} & $H$ & 18/20 & 0.196$\pm$0.047 & &$-$26.5$\pm$0.9 & $-$25.7$\pm$0.6\\
RQQ \citep{perc01}  & $K$ & 12/14 & 0.362$\pm$0.061 &$-25.6\pm0.8$ &$-$27.4$\pm$0.9 & $-$25.0$\pm$0.4\\
RQQ \citep{floy04} & $V$  & 10/10 & 0.390$\pm$0.031 & &$-$26.2$\pm$1.5 & $-$26.3$\pm$0.3 \\
RQQ \citep{hoop97} & $R$ & 10/10 & 0.433$\pm$0.032 & &$-$25.8$\pm$0.8 & $-$25.6$\pm$0.5\\
RQQ \citep{kuku01} & $J$  & 4/5  & 0.931$\pm$0.038 & &$-$25.4$\pm$0.9 & $-$26.1$\pm$0.5\\
RQQ \citep{falo04} & $H/K$ & 6/7 & 1.519$\pm$0.165 & $-26.7\pm0.8$ &$-$29.2$\pm$1.2 & $-$26.6$\pm$0.2 \\
RQQ \citep{peng06} & $H$ & 14/14 & 1.54$\pm$0.21 &                 &$-$27.9$\pm1.2$ & $-$26.1$\pm$0.8 \\
RQQ \citep{koti06} & $H/K$ & 6/6 & 1.56$\pm$0.25 &$-25.5\pm0.5$ & $-$27.3$\pm$0.5 & $-$26.6$\pm$0.9 \\
RQQ \citep{kuku01} & $H$ & 5/5 & 1.856$\pm0.120$ & & $-27.3\pm0.5$ & $-26.6\pm0.9$ \\
                   &           &                 &                 &                  \\
$L^*$ \citep{moba93} & $K$ & 136 & 0.077$\pm$0.030 & & & $-$25.0$\pm$0.2 \\
	     &	 &   &                 &               &             \\
BCG \citep{thua89} & $H$ & 84 & 0.074$\pm$0.026 & & &$-$26.3$\pm$0.3 \\
BCG \citep{arag98} & $K$ & 25 & 0.449$\pm$0.266 & & &$-$27.0$\pm$0.3\\
	&	&   &       &         &             \\
	&	&   &       &         &             \\
FSRQ \citep{koti98a} & $H$ & 9/16 & 0.671$\pm$0.157 &$-$26.2$\pm1.1$ &$-$29.7$\pm$0.8 & $-$26.7$\pm$1.2 \\
SSRQ \citep{koti00} & $H$ & 16/19 & 0.690$\pm$0.088 &$-$25.6$\pm1.0$ &$-$28.3$\pm$1.3 & $-$27.0$\pm$1.2 \\
\noalign{\smallskip}
\hline
\end{tabular}
\begin{list}{}{}
\item[$^{\mathrm{a}}$]
Column (1) gives the sample; (2) the filter; (3) resolved/total number of objects in the sample; 
(4) the average redshift of the sample; (5) the absolute $B$-band magnitude 
of the quasar (6) and (7) the 
average $H$-band nuclear and host galaxy absolute magnitude of the sample.
\item[$^{\mathrm{b}}$]
R = resolved; U = unresolved
\item[$^{\mathrm{c}}$]
Transformation of magnitudes to $H$-band assumes $V-H=3.0$, $R-H=2.5$, 
$H-K=0.2$ and $J-H=0.9$ for the hosts and $H-K=1$ for the nucleus. 
\end{list}
\end{table*}

In Table~\ref{samples}, we present a comparison of the average host properties 
of our RQQ sample those of other previous studies on RQQ, SSRQ and FSRQ hosts at similar and lower and higher redshift. 
In order to treat these data homogeneously, the apparent magnitudes reported in each literature study 
were transformed into absolute $H$-band magnitudes in our adopted cosmology, after $K$-correction 
and colour correction, assuming average rest-frame colours for giant ellipticals of $H-K=0.2$ \citep{reci90}, 
$R-K=2.7$ \citepalias{koti98a}, and $J-H=0.9$ and $V-H=3.0$ \citepalias{koti00}.

\begin{figure}
\centering
\resizebox{8cm}{15cm}{\includegraphics[bb=1.5cm 8.22cm 10.5cm 24.7cm,clip]{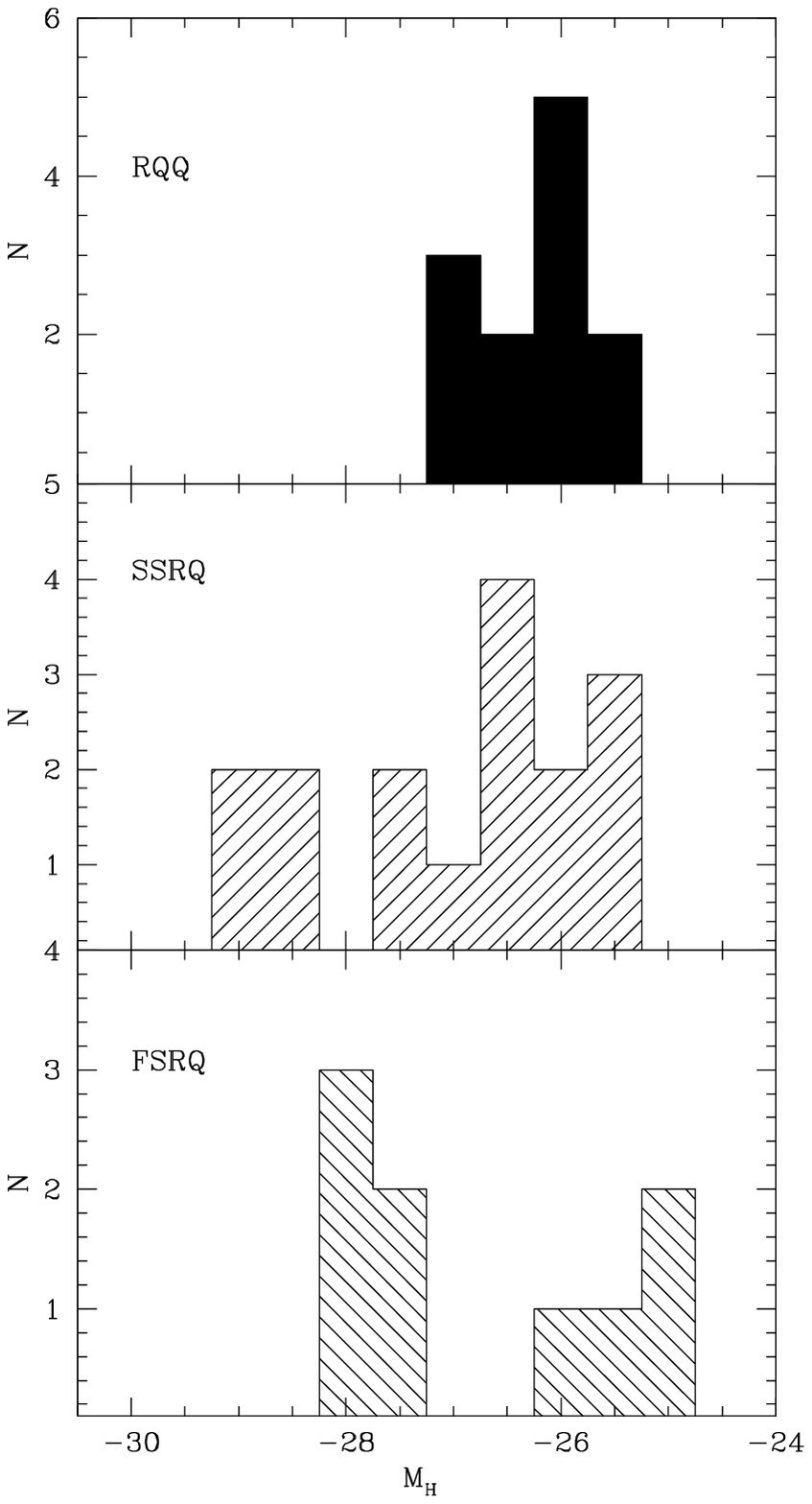}}
\caption[]{Histogram of the absolute $H$-band magnitude of the 12 RQQ hosts (top; this work), 
16 SSRQ hosts (middle; \citetalias{koti00}) and 9 FSRQ hosts (bottom; \citetalias{koti98a}) 
at the redshift range $0.5<z<1$.
\label{Mhhist}}
\end{figure}

According to the AGN unified models \citep{urry95}, FSRQs are strongly beamed objects, 
and in order to avoid any bias due to the contamination from beaming 
(by an unknown factor) of the nuclear luminosities of 
the FSRQs we compare the results for RQQs with mainly those for SSRQs. 
Fig.~\ref{Mhhist} shows the distribution of the RQQ, SSRQ and FSRQ host galaxy absolute $H$-band 
magnitudes (this work, \citetalias{koti00} and \citetalias{koti98a}, respectively) 
in the redshift range $0.5<z<1$. The average absolute magnitude of the 12 resolved RQQ hosts 
is $M_H=-26.3\pm0.6$, and including the three upper limits, $M_H^{all} = -26.2\pm0.6$ 
for the full sample of 15 RQQs. To estimate the total dispersion we used the equation for the variance 
and added that value to the dispersion of the resolved observations. 
All the observed RQQs have host galaxies with luminosity ranging between $M^*_H$ and $M^*_H-2$, where 
$M^*_H=-25.0$ \citep{moba93} is the characteristic luminosity of 
the Schechter luminosity function for elliptical galaxies. 
RQQ hosts belong, therefore, preferentially to the high luminosity tail of the galaxy luminosity function.
Note that the intermediate redshift RQQ hosts are on average $\sim1$ mag fainter than SSRQ hosts 
($M_H=-27.0\pm1.2$) in the same redshift range, even though the nuclear absolute magnitudes of 
the two samples are similar ($M_H=-27.9\pm1.0$ and $M_H=-28.3\pm1.3$ for RQQs and SSRQs, respectively). 
This difference is due to a tail of high luminosity SSRQ hosts ($M_H >$ -27) that are not present among 
the RQQ hosts. 
Similar difference between the luminosities of RQQ and RLQ hosts has been found in previous studies of 
quasars at both lower \citep{dunl03} and higher redshift \citep{kuku01,falo04,koti06}. 
For our intermediate redshift RQQ and RLQ samples, selection effects due to e.g. non-homogeneous 
distribution in redshift, optical luminosity and/or modeling of the host galaxy are irrelevant. 
Our results, therefore, strengthen the conclusion from previous studies that the gap in 
the host luminosity is intrinsic and remains the same over a wide redshift range between $z=0$ and $z=2$. 
Note that if the host luminosity is related to the bulge mass and thus to the BH mass, 
and on average RLQ hosts are brighter than RQQ hosts, the BHs in RLQs must be more massive 
than those in RQQs. Furthermore, to produce a similar optical luminosity, the BHs in RLQs must also be accreting 
less efficiently than the BHs in RQQs.

\begin{figure*}
\centering
\includegraphics[width=15cm]{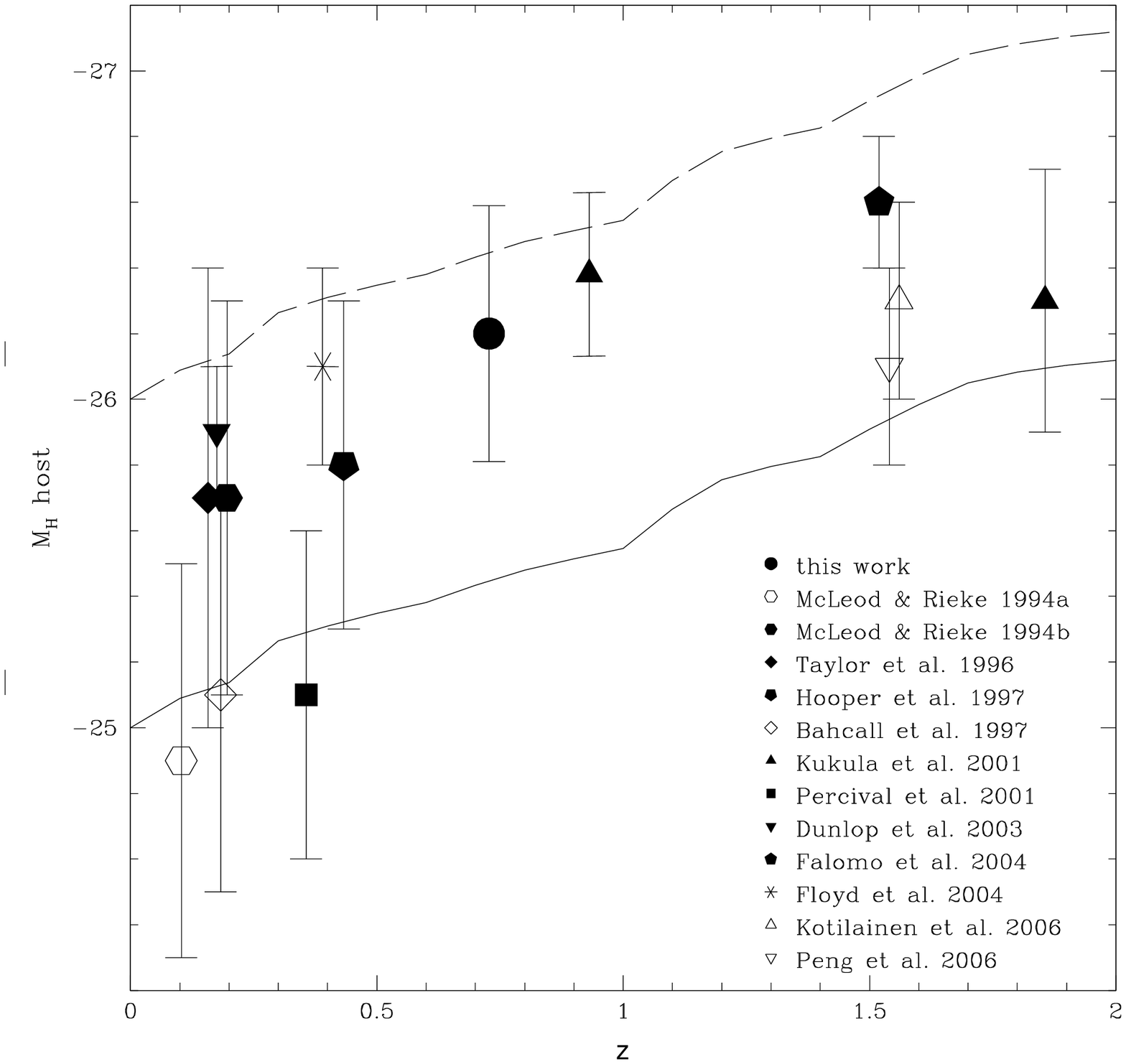}
\caption{The average absolute $H$-band magnitude of RQQ host galaxies 
as a function of redshift. The combined sample of resolved and unresolved RQQs (this work) is marked as 
filled circle, sample from \citet{kuku01} as triangles, \citet{dunl03} as inverted triangle, 
\citet{perc01} as square, \citet{falo04} as pentagon, \citet{floy04} as asterisk, 
\citet{mcle94a,mcle94b} as open and filled hexagon, \citet{tayl96} as diamond, \citet{hoop97} 
as inverted pentagon, \citet{bahc97} as open diamond, \citet{koti06} as open triangle and 
\citep{peng06} as inverted open triangle. The solid and 
long-dashed lines are the luminosities of $L^*$ ($M_H=-25.0$ at low redshift; \citet{moba93}) and 
$L^*-1$ galaxies following the passive evolution model of \citet{bres98}.
\label{Mhevol_tot_rqq}}
\end{figure*}

Fig.~\ref{Mhevol_tot_rqq} shows the average $H$-band absolute magnitudes 
derived from the various samples of RQQ host galaxies (Table~\ref{samples}) as a function of redshift, 
based on NIR HST and ground-based data (for explanation of symbols, see the caption of 
Fig.~\ref{Mhevol_tot_rqq}). All these data have been made consistent with 
our system (as regards extinction, $K$-correction and cosmology) starting from 
the total apparent magnitudes of the host galaxies. 

Note that at $z<0.5$ the scatter of the average host galaxy magnitudes between the various samples is 
much larger than that at higher redshift. Notably, the average host galaxy magnitudes of 
the low redshift RQQs observed by \citet{mcle94a}, \citet{bahc97}, and \citet{perc01} are $\sim$0.5-1 mag 
fainter than those of the majority of the low redshift RQQ samples. 
If there is a correlation between nuclear and host luminosities (see next Section), 
the low luminosity RQQs observed by \citet{mcle94a} (open hexagon) are indeed expected to have 
relatively low luminosity hosts, confirmed by comparing with their high luminosity RQQ sample 
\citep{mcle94b} (filled hexagon). 
The case of the discrepant value from the high nuclear luminosity RQQ sample of \citet{perc01} is more 
difficult to interpret. In the host galaxy fitting, they used a varying $\beta$ parameter, including several 
cases with a pure exponential disk model, leading to significantly lower host magnitudes than those with 
an elliptical galaxy model. Furthermore, from their reported data it is not possible to ascertain 
the data quality (e.g. no luminosity profiles were shown). 

Overall, the intermediate redshift RQQ hosts follow the same trend as RQQ hosts at lower and higher redshift. 
They are consistent with a passive stellar population evolution of massive ellipticals, 
in the redshift range $0<z<2$. This scenario of a passive evolution of quasar hosts agrees with 
the few available spectroscopic studies of low redshift quasar hosts \citep{cana00,nola01}, 
indicating that their stellar content is dominated by an old evolved stellar population. 
The cosmic evolution traced by quasar hosts up to z $\sim$2 disagrees with semianalytic 
hierarchical models of AGN and galaxy formation and evolution \citep{kauff00}, 
which predict fainter (less massive) hosts at high redshift, which merge and grow to form the low redshift 
massive spheroids. Thus, if quasar hosts are luminous spheroids undergoing passive evolution, 
their mass remains essentially unchanged from z $\sim$2 up to the present epoch. 

The average effective radius of the 12 resolved intermediate redshift RQQ hosts is 
$R_e=11.3\pm5.8$ kpc. This is in good agreement with intermediate redshift SSRQ and FSRQ hosts 
for which $R_e=8.6\pm1.9$ and $R_e=12.8\pm6.0$ kpc, (\citetalias{koti00,koti98a}, respectively). 
The quoted uncertainties are due to the dispersion of the distribution, 
while the individual large dispersions due to the degeneracy between the effective radius R$_e$ and 
the surface brightness $\mu_e$ have not been taken into account. 
The intermediate redshift RQQ hosts have also similar sizes to those of both lower redshift RQQ hosts, e.g. 
$R_e=11.4\pm1.7$ kpc \citep{dunl03} and $R_e=8.7\pm1.8$ kpc \citep{floy04}, 
and to those of higher redshift RQQ hosts, e.g. $R_e=11.6\pm2.5$ kpc \citep{falo04} 
and $R_e=6.5\pm1.6$ kpc \citep{koti06}. 
This confirms that the effective radius of RQQ hosts does not evolve with redshift, and suggests 
that at redshift $z>0.5$ the RQQ hosts have similar dynamical structure of normal (presently inactive) 
elliptical galaxies.

\subsection{Correlation between nuclear and host luminosity}

The absolute magnitudes of the nuclear component in the intermediate redshift RQQs range from 
$M_H \sim-26$ to $M_H \sim-29.5$, with an average value of $M_H=-27.9\pm1.0$, 
in good agreement with that of the average nuclear magnitude of SSRQs ($M_H=-28.3\pm1.3$; \citetalias{koti00}). 
On the other hand, the nuclear component of the FSRQs is much brighter, $M_H=-29.7\pm0.8$ 
(\citetalias{koti98a}), consistent with their nuclear emission being boosted by relativistic beaming. 

If the mass of the central BH is proportional to the luminosity of the spheroid of the host galaxy, 
as observed for nearby massive early-type galaxies, and if the quasar is emitting at 
a fixed fraction of the Eddington luminosity, one would expect a correlation between 
the luminosity of the nucleus and that of the host galaxy. 
However, nuclear obscuration, beaming, and/or an intrinsic spread in 
accretion rate and mass-to-luminosity conversion efficiency, may destroy 
this correlation. Our combined sample (this work; \citetalias{koti00} and \citetalias{koti98a} covers 
a broad range of nuclear luminosity ($-24<M_B<-27$) and can therefore be used to investigate this issue.

\begin{figure*}
\centering
\includegraphics[width=15cm]{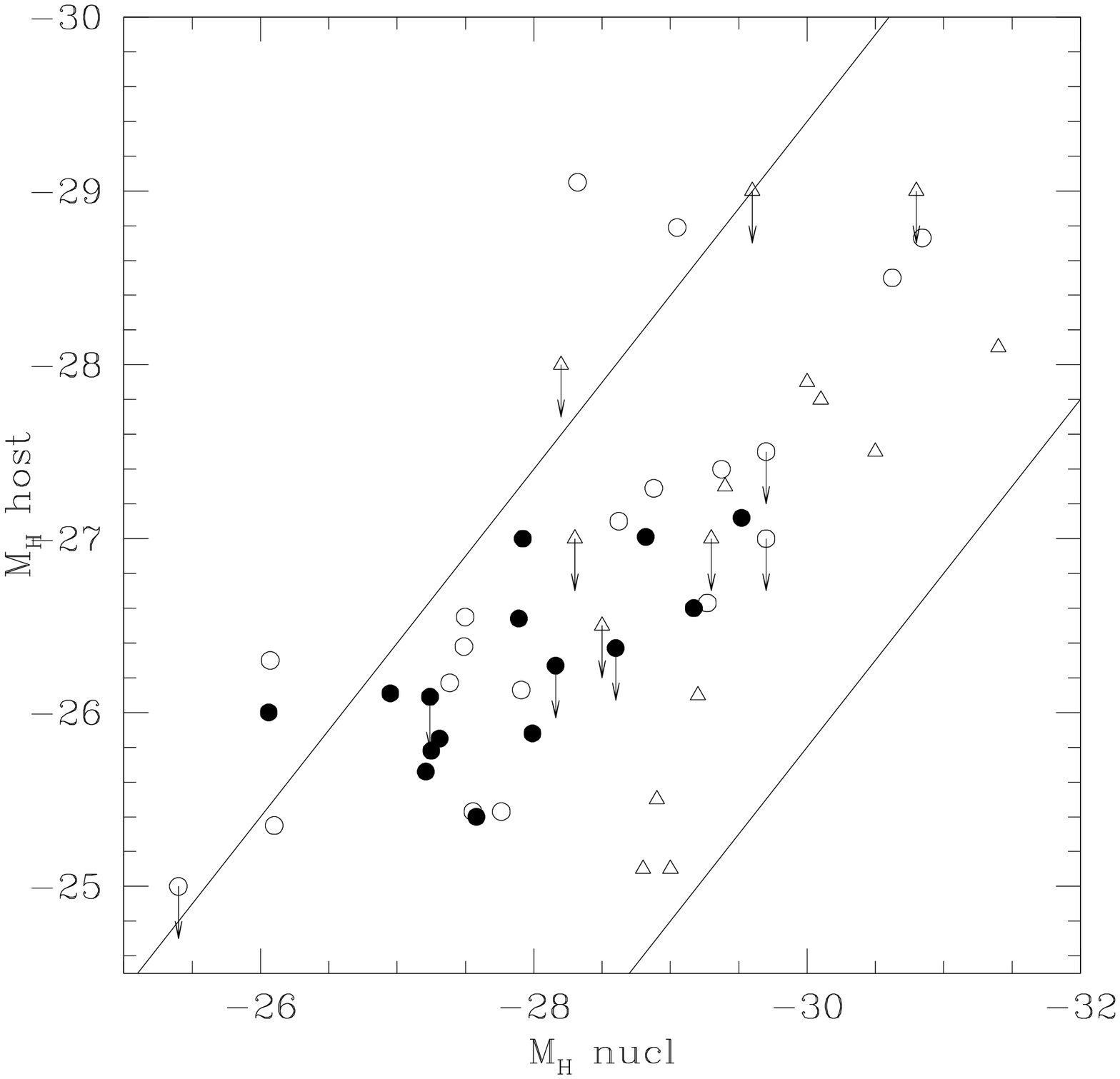}
\caption{The $H$-band absolute magnitude of the nucleus compared with that of the host galaxy. 
RQQs from this work are marked as filled circles. while SSRQs from \citetalias{koti00} 
and FSRQs from \citet{koti98a} are marked as open circles and open triangles, respectively. 
The arrows represent the upper limits of the host luminosity for unresolved objects. 
The diagonal lines represent the loci of constant ratio between host and nuclear emission. 
These can be translated into Eddington ratios assuming that the central black hole mass - galaxy luminosity 
correlation holds at high redshifts and that the observed nuclear power is proportional to 
the bolometric emission. The diagonal lines encompass a spread of 1.5 dex in the nucleus/host luminosity ratio. 
\label{MhMn_new}}
\end{figure*}

Fig.~\ref{MhMn_new} shows the $H$-band absolute host galaxy magnitude versus the absolute nuclear magnitude 
for intermediate redshift RQQs (this work), FSRQs and SSRQs (\citetalias{koti98a,koti00}, respectively). 
As noted in the previous Section, the intermediate redshift RQQs tend to occupy a region of less luminous 
host galaxies than the SSRQs. 
However, before any conclusions are drawn it is worth noting the obvious selection effects 
that could introduce a spurious apparent correlation between the nuclear and host luminosities. 
In particular, two such effects may combine to depopulate the $M_{nuc}-M_{host}$ plane in opposite 
directions. Firstly, a faint host would be very difficult to detect against a bright nucleus 
(upper left hand region). Secondly, a low luminosity nucleus would be difficult to detect against 
a bright host galaxy (lower right hand region). In our sample, the first effect should not be 
very serious because only three RQQs remained unresolved and their upper limits do not populate 
an extreme region of the diagram. The second effect is likely to be small because of the rareness of 
extremely luminous galaxies.

We find a reasonably strong nuclear luminosity dependence of the host galaxy luminosity 
for the full sample of quasars (RQQs and SSRQs), with a Spearman rank correlation coefficient $R_S=0.737$. 
This correlation is present also considering the RQQs and SSRQs separately 
($R_S=0.606$ and $R_S=0.713$, respectively). 
The RLQs and RQQs in our study have a similar distribution in their nuclear luminosity, 
and thus we believe this to be a robust result. 
Note that we have found a similar dependence at even higher redshift for RLQs, and to a lesser extent for RQQs 
\citep{koti06}. Since generally no such correlation has been found at low redshift \citep[e.g.][]{dunl03,paga03}, 
(but see e.g. \citet{sanc04} who found a correlation for low luminosity AGN), 
it must have its onset at a relatively high redshift. 
Note that the correlation suggests that the low redshift relation between 
$M_{BH}$ and $M_{bulge}$ holds also at high redshift, but its confirmation requires 
a spectroscopic determination of BH masses in high redshift quasars. 
Note also that there is a large scatter in the relationship, possibly due to 
varying accretion efficiency, and intrinsic scatter in the $M_{BH}-M_{bulge}$ relation. 

\begin{figure}
\centering
\resizebox{8cm}{15cm}{\includegraphics[bb=1.5cm 8.22cm 10.5cm 24.7cm,clip]{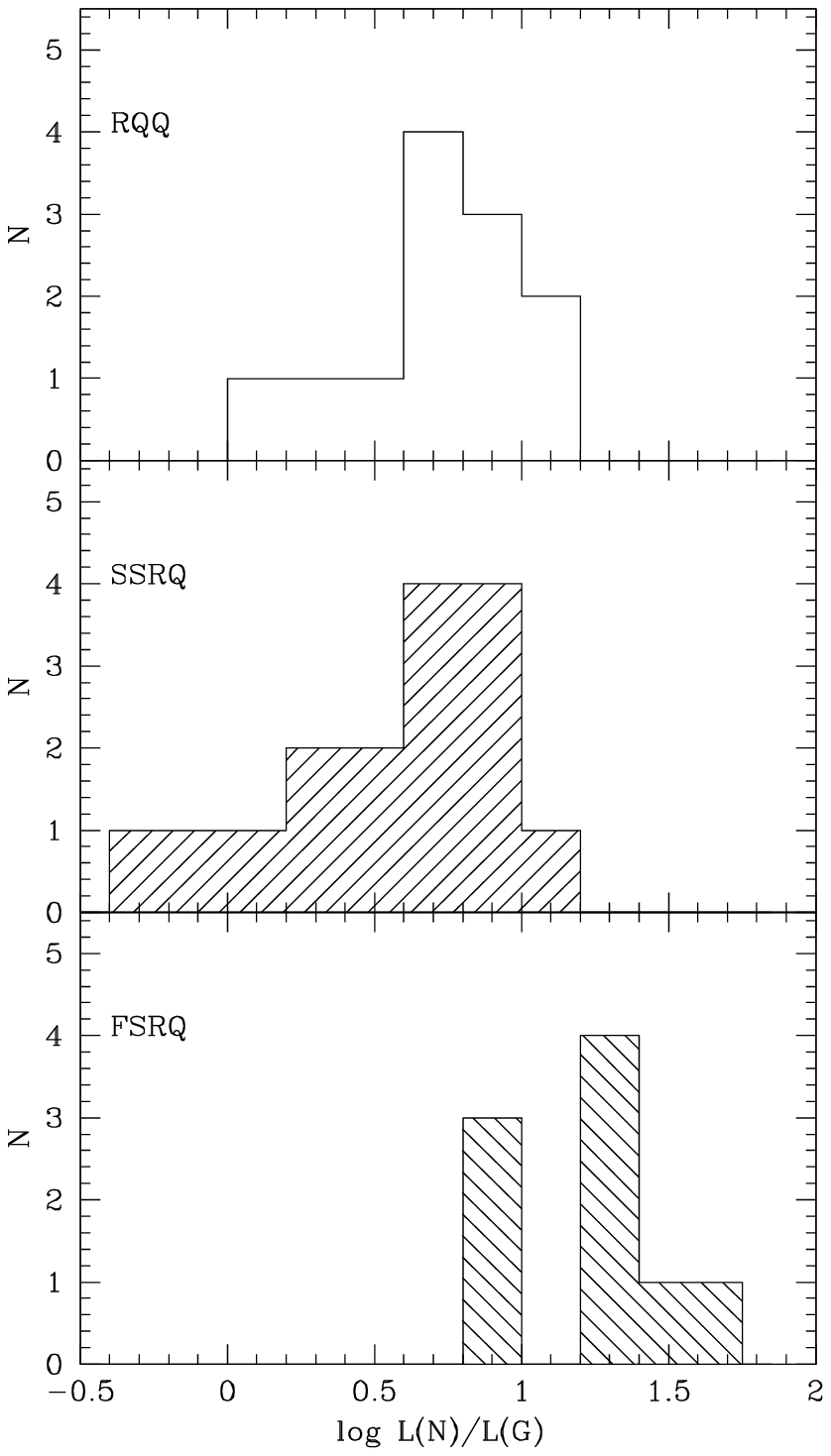}}
\caption[]{Histogram of the nucleus/host luminosity ratio for the resolved RQQs (top), 
SSRQs (middle; \citetalias{koti00}) and FSRQs (bottom; \citetalias{koti98a}).
\label{histos}}
\end{figure}

Assuming that the the $H$-band nuclear luminosity is proportional to the bolometric luminosity and that 
the host galaxy luminosity is proportional to the BH mass (as observed at low redshift), 
the nucleus/host luminosity ratio should be proportional to the Eddington factor $L/L_E$, 
where $L_E=1.25\times10^{38}\times(M_{BH}/M_{\odot})$. 
Fig.~\ref{histos} shows the distribution of the $H$-band nucleus/host luminosity ratio 
for the samples of resolved RQQs, FSRQs and SSRQs. 
The spread in the nucleus/host luminosity ratio for RQQs is similar to that of SSRQs but smaller than 
that of FSRQs. For the resolved intermediate redshift RQQs, the average $H$-band nucleus/host ratio is 
$log(L_{nuc}/L_{host}) = 0.71\pm0.29$, 
slightly higher than that in intermediate redshift SSRQs $(log(L_{nuc}/L_{host}) = 0.58\pm0.51)$, 
but lower than that in the FSRQs ($log(L_{nuc}/L_{host})=1.32\pm1.04$), consistent with 
the idea that the nuclear luminosities of the FSRQs are affected by strong nuclear beaming. 
The nucleus/host ratio of intermediate redshift RQQs is 
lower than that found for  higher redshift high luminosity RQQs with 
$log(L_{nuc}/L_{host})=1.32\pm0.37$ \citep{falo04}, 
but is consistent with the ratio observed for high redshift low luminosity RQQs with 
$log(L_{nuc}/L_{host})=1.00\pm0.44$ \citep{koti06} and $log(L_{nuc}/L_{host})=0.61\pm0.69$ \citep{kuku01}. 
Thus, $L/L_E$ appears to depend on quasar luminosity, being significantly higher in 
high luminosity quasars (see \citet{koti06}). This may indicate a smaller accretion efficiency 
in low luminosity quasars.

\begin{table}
\centering
\caption{The black hole masses of the RQQs.$^{\mathrm{a}}$}
\label{bhmass}
\begin{tabular}{lc}
\hline
\noalign{\smallskip}
Object & $M_{BH}/10^{9}M_{\odot}$\\
\noalign{\smallskip}
\hline
\noalign{\smallskip}
HS 0010+3611   & 0.5 \\
1E 0112+3256   & 1.3 \\
PB 6708        & $<$0.7 \\
KUV 03086-0447 & 2.0 \\
US 3828        & 0.5 \\
MS 08287+6614  & 0.7 \\
TON 392        & 1.3 \\
US 971         & 0.4 \\
HE 0955-0009   & $<$0.9 \\
HE 1100-1109   & 2.2 \\
CSO 769        & 1.1 \\
HS 1623+7313   & 0.3 \\
HS 2138+1313   & $<$1.0 \\
LBQS 2249-0154 & 1.2 \\
ZC 2351+010B   & 0.5 \\
\noalign{\smallskip}
\hline
\end{tabular}
\begin{list}{}{}
\item[$^{\mathrm{a}}$]
Transformations of magnitudes to $R$-band done assuming $R-H = 2.5$.
\end{list}
\end{table}

Assuming that the $R$-band host galaxy luminosity is proportional to the BH mass, 
we can estimate the BH mass for each resolved RQQ using the relation 
$log(M_{bh}/M_{\odot})=-0.50(\pm0.02)M_R - 2.96(\pm0.48)$ \citep{mclu02,marc03}. 
We have used $R-K=2.7$ \citepalias{koti98a} and $H-K=0.2$ \citep{reci90} to transform 
the $H$-band magnitudes to the $R$-band. 
The resulting BH masses are given in Table~\ref{bhmass}. The average value for 
the BH masses of RQQs is $M_{BH,RQQ}=1.0\pm0.6\times10^9$ $M_{\odot}$, with five out of the 12 RQQs 
having BH masses above $10^{9}$ $M_{\odot}$. This is consistent with the average BH mass 
for low redshift RQQs, $M_{BH,RQQ}=1.05\pm0.24\times10^9$ $M_{\odot}$ \citet{dunl03}. 
These values can be compared with the BH masses of the 
intermediate redshift RLQs, $M_{BH,RLQ}=4.1\pm5.6\times10^9$ $M_{\odot}$ \citepalias{koti98a,koti00}. 
The BH masses of the RLQs thus seem to be about four times larger than the BH masses of the RQQs. 
This may reflect a fundamental difference between RQQs and RLQs and supports the idea that 
the radio properties of quasars depend on the mass of the central BH.

\subsection{The close environments of RQQs}

Quasars often have close companions and their host galaxies sometimes exhibit a disturbed morphology 
\citep[e.g.][]{yee84,stoc87,hutc90,hutc95,bahc97,hutc99}.
Spectroscopic studies \citep[e.g.][]{heck84,cana97} have shown 
that indeed in many cases the companions are at the redshift of the quasar and are 
therefore physically associated. This is consistent with the long standing idea that strong tidal 
interactions and galaxy mergers can trigger and/or fuel the nuclear activity.

Both RLQs and RQQs seem to inhabit relatively dense environments (e.g. groups of galaxies) but are only rarely 
located in rich galaxy clusters. Although some earlier studies suggested that RQQs are generally found 
in poorer environments than RLQs \citep[e.g.][]{elli91}, the majority of recent studies 
\citep[e.g.][]{wold00,wold01,mclu01} have come to the conclusion that there is 
no significant difference between the large scale environments of RQQs and RLQs. 
Clarifying this issue is important for understanding the connection between the characteristics of 
the nuclear activity (e.g. the level of nuclear radio emission) and the environment.

\begin{table}
\centering
\caption{The close environments of the RQQs.$^{\mathrm{a}}$}
\label{environment}
\begin{tabular}{@{}llll@{}}
\hline
\noalign{\smallskip}
Name & $z$ & $N_r$ & $N_r$ \\
 & & (50 kpc) & (100 kpc) \\
\noalign{\smallskip}
\hline
\noalign{\smallskip}
HS 0010+3611   & 0.530 & 0 & 0\\
1E 0112+3256   & 0.764 & 0 & 1\\
PB 6708        & 0.868 & 1 & 1\\
KUV 03086-0447 & 0.755 & 0 & 0\\
US 3828        & 0.515 & 1 & 1\\
MS 08287+6614  & 0.610 & 0 & 1\\
TON 392        & 0.654 & 0 & 1\\
US 971         & 0.703 & 0 & 0\\
HE 0955-0009   & 0.597 & 0 & 0\\
HE 1100-1109   & 0.994 & 0 & 3\\
CSO 769        & 0.850 & 0 & 0\\
HS 1623+7313   & 0.621 & 0 & 0\\
HS 2138+1313   & 0.810 & 2 & 3\\
LBQS 2249-0154 & 0.832 & 0 & 2\\
ZC 2351+010B   & 0.810 & 0 & 1\\
\noalign{\smallskip}
\hline
\end{tabular}
\begin{list}{}{}
\item[$^{\mathrm{a}}$]
Column (1) gives the name of the object; (2) the redshift of the object; (3) the number of 
companions within 50 kpc radius; (4) the number of companions within 100 kpc radius.
\end{list}
\end{table}

Following the method in \citetalias{koti00}, 
we have compared the frequency of companions found around the samples of RQQs and RLQs 
at intermediate  $z$ discussed in the previous sections. 
For each RQQ, we counted the number of resolved companion objects, i.e. galaxies, within a specified radius 
that are brighter than $m^*_H + 2$, where $m^*_H$ is the apparent magnitude corresponding to $M^*_H=-25$ at 
the redshift of each quasar. This limit is in all cases at least 1 mag brighter than the magnitude limit of 
the images. We selected two radii corresponding to projected distances of 50 kpc and 100 kpc around the quasar. 
In our adopted cosmology, 50 kpc corresponds to $\sim5.3$ arcsec and 100 kpc to $\sim10.6$ arcsec at 
the average redshift of the samples, $z\sim$0.7. 
In Table~\ref{environment} we give the number of resolved companions ($N_r$) within 50 and 100 kpc 
projected distance from each RQQ.

\setlength{\tabcolsep}{1.5mm}
\begin{table}
\centering
\caption{Close companions of RQQs and RLQs.$^{\mathrm{a}}$}
\label{average}
\begin{tabular}{@{}lllll@{}}
\hline
\noalign{\smallskip}
Sample & $N_{ave}$ & $N_{ave}$ & $N_{c}/N_{tot}$ & $N_{c}/N_{tot}$ \\
 & (50 kpc) & (100 kpc) & (50 kpc) & (100 kpc) \\
\noalign{\smallskip}
\hline
\noalign{\smallskip}
RQQ & 0.27$\pm$0.59 & 0.93$\pm$1.03 & 20\% & 60\%\\
(this work) & & & & \\
FSRQ & 0.19$\pm$0.39 & 1.19$\pm$1.01 & 19\% & 75\%\\
\citepalias{koti98a} & & & & \\
SSRQ & 0.05$\pm$0.22 & 0.53$\pm$0.68 & 5\%  & 42\%\\
\citepalias{koti00} & & & &\\
\noalign{\smallskip}
\hline
\end{tabular}
\begin{list}{}{}
\item[$^{\mathrm{a}}$]
Column (1) gives the sample; (2) the average number of companions within 50 kpc radius; 
(3) the average number of companions within 100 kpc radius; (4) and (5) the fraction of quasars 
with close companion within 50 kpc and 100 kpc radius, respectively.
\end{list}
\end{table}

The average number of companions around the RQQs (this work), FSRQs and SSRQs 
(\citetalias{koti98a,koti00}, respectively) are reported in Table~\ref{average}, 
together with the fraction of quasars that have at least one close companion. 
RQQs appears to inhabit environments similar to FSRQs but richer than SSRQs. 
It seems that there is not a direct link with radio propertes between RQQs and RLQs which 
consistent with the results by \citet{smit00,wold01,mclu01}. From our results it appears 
that neither RQQs and RLQs have significant numbers of close companions to be important 
for the triggering and fuelling of nuclear activity. However, it remains a possibility 
that the timescales of the interaction and the triggering are so different 
that by the time of the onset of the nuclear activity, no detectable sign of the past interaction remains.

\section{Conclusions}

We have presented a  near-infrared imaging study of a 
sample of radio-quiet quasars at $0.5<z<1$. In 12 out of the 15 quasars observed, we were able to resolve 
the objects and to characterize the properties of their host galaxies. 
The RQQ host galaxies at $z\sim0.7$ are luminous elliptical galaxies, with average magnitude $M_H=-26.3\pm0.6$, 
$\sim1$ mag brighter than low redshift $L^*$ galaxies. Our results are consistent with the passively 
evolving stellar population that was created at high redshift ($z\sim3-4$). 
RQQ hosts are $\sim0.5$ magnitude fainter than those of RLQs at all redshifts in the range $0<z<2$.
The host galaxies of the intermediate redshift RQQs are giant ellipticals, with an 
average effective radius $R_e=11.3\pm5.8$ kpc. This is similar to 
those of lower and higher redshift RQQs, indicating that the effective radius of the host galaxies 
does not evolve with redshift. The intermediate redshift 
RQQs tend to have fainter nuclear luminosity than FSRQs and SSRQs at the same redshift, suggesting that their 
central engine is less powerful. RQQs have relatively rich environments, similar to the environments of FSRQs. 

 
\begin{acknowledgements}
TH, JKK and E\"O acknowledge financial support from the Academy of Finland, 
projects 8107775 and 8201017. This work has been partially supported by INAF contract 1301/01.
\end{acknowledgements}

\bibliographystyle{aa} 
\bibliography{totahy.bib}

\end{document}